\documentclass[superscriptaddress,twocolumn,showpacs,amsmath,amssymb,pre]{revtex4}

\usepackage{graphicx}
\usepackage{dcolumn}
\usepackage{bm}
\usepackage{bbm}
\usepackage{multirow,amssymb,amsbsy,amsmath}
\usepackage{stmaryrd}

\newcommand{\I}{i}
\newcommand{\E}{e}
\newcommand{\D}{d}

\newcommand{\dd}{d}
\newcommand{\dod}[2]{\frac{\dd #1}{\dd #2}}

\newcommand{\ddim}{\udelta\kern0.1em}

\newcommand{\beikonst}[2]{\left( #1 \right)_{\kern-0.2em #2}}

\newcommand{\trtxt}[2][]{\text{Tr}_{#1}\{#2\}}
\newcommand*{\bra}[1]{\mathopen{\langle}#1\mathclose{|}}
\newcommand*{\ket}[1]{\mathopen{|}#1\mathclose{\rangle}}

\newcommand{\comutxt}[2]{[#1,#2]}

\newcommand{\h}[1]{\hat{H}_{#1}}
\newcommand{\dop}{\hat{\rho}}
\newcommand{\PI}[1]{\hat{\Pi}_{#1}}

\begin{document}

\preprint{APS/123-QED}

%
%
\title{Transport in anisotropic model systems analyzed by a correlated projection superoperator technique}

\author{Hendrik Weimer}%
\affiliation{Institute of Theoretical Physics I, University of
  Stuttgart, Pfaffenwaldring 57, 70550 Stuttgart, Germany}%
\email{hweimer@itp1.uni-stuttgart.de}%
\author{Mathias Michel}%
\affiliation{Advanced Technology Institute, Faculty of Engineering and Physical Sciences, University of Surrey, Guildford GU2 7XH, United Kingdom}%
\author{Jochen Gemmer}%
\affiliation{Physics Department, University of Osnabr\"uck,
  Barbarastr.\ 7, 49069 Osnabr\"uck, Germany}%
\author{G\"unter Mahler}%
\affiliation{Institute of Theoretical Physics I, University of
  Stuttgart, %
  Pfaffenwaldring 57, 70550 Stuttgart, Germany}%

\date{\today}%

\begin{abstract}
  By using a correlated projection operator, the time-convolutionless
  (TCL) method to derive a quantum master equation can be utilized to
  investigate the transport behavior of quantum systems as well.
  Here, we analyze a three-dimensional anisotropic quantum model
  system according to this technique.  The system consists of
  Heisenberg coupled two-level systems in one direction and weak
  random interactions in all other ones. Depending on the partition
  chosen, we obtain ballistic behavior along the chains and normal
  transport in the perpendicular direction.  These results are
  perfectly confirmed by the numerical solution of the full
  time-dependent Schr\"odinger equation.
\end{abstract}



\maketitle

%
%

%
\section{Introduction}
\label{sec:1}

The transport of different extensive quantities like energy, heat,
entropy, mass, charge, magnetization, etc., through and within solid
state systems is an intensively studied topic of nonequilibrium
statistical dynamics.  Nevertheless, there are numerous open questions
concerning the type of transport especially in small systems far from
the thermodynamic limit and in particular in quantum mechanics.  At
the heart of many investigations is the classification into two main
categories: \emph{normal} or \emph{diffusive} transfer of the
extensive quantity and \emph{ballistic} transport featuring a
divergence of the conductivity.

Diffusive transport occurs whenever the system is governed by a
diffusion equation. In particular, this means that excitations decay
exponentially fast and the spatial variance of an initial excitation
grows linear in time. Ballistic transport, however, is rather
described by the equations of motion of a free particle. For the
spatial variance of an excitation this implies a quadratic growth in
time.


In the present paper, we will concentrate on the transport of energy
and heat in quantum systems.  There are several different approaches
discussed in the literature to investigate the transport of those
quantities in quantum mechanics.  One very famous ansatz is the
investigation of heat transport in terms of the Green-Kubo formula
\cite{Zotos1997,Prosen1999,Kluemper2002,Heidrich-Meisner2003,Saito2003,Jung2006}.
A main advantage of this approach is certainly its computability after
having diagonalized the system's Hamiltonian.  Derived on the basis of
linear response theory the Kubo formula has originally been formulated
for electrical transport \cite{Kubo1957,Kubo1991}, where an external
potential can be written as an addend to the Hamiltonian of the
system.  Basically one finds a current-current autocorrelation, which
has \emph{ad hoc} been transferred to heat transport simply by
replacing the electrical current by a heat current
\cite{Luttinger1964}.  However, the justification of this replacement
remains unclear since there is no way of expressing a temperature
gradient in terms of an addend to the Hamiltonian of the system as
before \cite{Gemmer2006}.

Other approaches to heat conductivity in quantum systems are based on
diagonalization of the Schr\"odinger equation \cite{Gobert2005},
analyzing the level statistics of the Hamiltonian
\cite{Mejia-Monasterio2005, Steinigeweg2006} or by an explicit
coupling to some environments of different temperature
\cite{Saito2003a, Michel2003}.  In the latter case, environments are
described by a quantum master equation \cite{Breuer2002} in Liouville
space.  Here the temperature differences can, indeed, be described by
a perturbation operator so that one may treat a thermal perturbation
in this extended state space similar as an electrical one in the
Hilbert space \cite{Michel2004}.

The Hilbert space Average Method \cite{Gemmer2004} allows for a direct
investigation of the heat transport in quantum systems from
Schr\"odinger dynamics. By deriving a reduced dynamical equation for a
class of design quantum systems, normal heat transport as well as
Fourier's Law has been confirmed \cite{Michel2005,Michel2006}.
Recently, it has been shown that for diffusive systems the Hilbert
space average method is equivalent to a projection operator technique
with an extended projection superoperator
\cite{Breuer2006,Breuer2007}. However, ballistic behavior cannot be
analyzed with the Hilbert space Average Method in a straightforward
manner since it is not obvious how to obtain time-dependent rates.

Using a correlated superprojection operator within the derivation of
the time-convolutionless (TCL) quantum master equation leads to a
reduced dynamical description of the investigated system.  The main
advantage of the correlated TCL method refers to its perturbation
theoretical character.  Thus it is a systematic expansion in some
perturbational parameter.

To use this alternative method for an investigation of the transport
behavior of a quantum system, it is necessary to partition the
microscopic system described by the Hamiltonian $\hat{H}$ into
mesoscopic subunits.  While the complete dynamics is governed by the
Schr\"odinger equation of the full system according to its density
operator
\begin{equation}
  \label{vng}
  \dot{\dop}=-\frac{\I}{\hbar}[\hat{H},\hat{\rho}]\equiv\mathcal{L}(t)\hat{\rho},
\end{equation}
we aim at deriving a closed reduced dynamical equation for the
subunits chosen. Formally, this partitioning is done by introducing a
projection superoperator $\mathcal{P}$ that projects onto the relevant
part of the full density matrix $\dop$ \cite{Breuer2002}, here the
spatial energy distribution within the system. However, by a
straightforward application of the projection superoperator on the
above equation, the dynamics of the reduced system is no longer
unitary, but described by
\begin{equation}
  \label{eq:vng2}
  \mathcal{P}\dot{\dop} = \mathcal{P}\mathcal{L}(t)\dop.
\end{equation}

These effective equations of motion for the relevant part
$\mathcal{P}\dop$ can either be written as an integro-differential
equation (Nakajima-Zwanzig equation \cite{Nakajima1958, Zwanzig1960})
or as a time-convolutionless (TCL) master equation \cite{Breuer2002},
which is an ordinary linear differential equation of first order. Both
methods allow for a systematic perturbative expansion.  In the TCL
expansion series the first-order term typically vanishes and thus the
leading order is given by (cf.~\cite{Breuer2006,Breuer2007})
\begin{equation}
  \label{eq:tcl}
  \mathcal{P}\dot{\dop} = -\int\limits_0^t \D t_1 \mathcal{P}\mathcal{L}(t)\mathcal{L}(t_1)\mathcal{P}\dop.
\end{equation}
However, in order to obtain a converging perturbation series expansion
$\mathcal{P}$ should not be chosen arbitrarily: A ``wrong'' projection
superoperator may lead to a breakdown of the expansion
\cite{Breuer2006}.



%
%
\section{Description of the model}
\label{sec:2}

In the present paper we consider a three-dimensional (3D) model
composed of two-level systems.  The coupling between the atoms is
anisotropic, i.e., in one direction dominated by a Heisenberg
interaction whereas the coupling in all other directions is
random. The choice of random couplings ensures that the interaction is
unbiased as it does not have any special symmetry.  Two-level atoms or
spin-$1/2$ systems \cite{Schollwock2004} allow to study a large
variety of quantum effects from quantum information processing to
solid state theory, described by a rather simple interaction, making
them interesting both from an experimental and theoretical point of
view.  Of particular interest are the transport properties of systems
containing 1D and 2D spin structures, e.g., the investigations of heat
transport in cuprates, in which a dramatic heat transport anisotropy
has been reported \cite{Sologubenko2000,Hess2001}. While the
anisotropy is mainly attributed to anisotropic phonon scattering
processes, we investigate transport anisotropies emerging from an
anisotropic (but coherent) interaction.

\begin{figure}
  \centering 
\includegraphics[width=7.5cm]{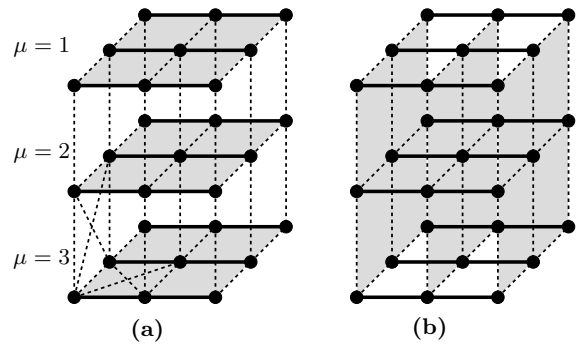}

  \caption{Partition schemes for investigating the transport
    perpendicular (a) or parallel (b) to the spin chains. Each spin is
    represented by a dot, solid lines indicate Heisenberg interactions
    along the chains, dashed lines represent random interactions.  The
    diagonal couplings within each plane have been left out for
    clarity [except the lower left corner of (a)].}
\label{fig:3d}
\end{figure}
The model we are going to investigate is a three-dimensional model of
two-level systems depicted in Fig.~\ref{fig:3d}.  In terms of Pauli
operators the local Hamiltonian of the network is given by
\begin{equation}
 	\label{eq:HZ}
 	\h{loc} = \frac{\Delta E}{2}\sum\limits_i\hat{\sigma}_z^{(i)},
\end{equation}
with the local energy splitting $\Delta E$ defining the basic energy
unit within our model.

In $x$ direction, the two-level systems are coupled via a Heisenberg
interaction
\begin{equation}
	\label{eq:HH}
	\h{H} = \sum\limits_i \hat{\boldsymbol{\sigma}}^{(i)}\otimes\hat{\boldsymbol{\sigma}}^{(i+1)},
\end{equation}
with the Pauli spin vectors $\hat{\boldsymbol{\sigma}}^{(i)} =
(\hat{\sigma}_x^{(i)}, \hat{\sigma}_y^{(i)}, \hat{\sigma}_z^{(i)})$ at
site $i$.

In the $y$ and $z$ directions, we use a random interaction matrix
$\h{R}$ to couple both adjacent sites and next neighbor sites lying
diagonally opposite [see lower left corner of Fig.~\ref{fig:3d}(a)].
The nonzero matrix elements are taken from a Gaussian ensemble with
zero mean and a variance $s^2$. While each matrix element is taken
from the same ensemble, the geometry of the system requires that we do
not have translational invariance within the random interaction.

To investigate the transport in $x$ or $z$ direction, respectively
(cf.~Fig.~\ref{fig:3d}), we perform a partition of the model into $N$
subunits.  A layer of $n$ two-level systems is grouped together into a
new local subsystem, coupled to adjacent layers by the connections
between pairs of two level systems.  Because of the anisotropy within
the model we can study the transport perpendicular to the Heisenberg
chains in the $z$ direction [Fig.~\ref{fig:3d}(a)] and along the
chains in $x$ direction [Fig.~\ref{fig:3d}(b)].

The coupling strength of an arbitrary interaction matrix $\hat{V}$ is
defined as
\begin{equation}
	\label{eq:mm1}
	\eta = \frac{1}{d}\sqrt{\trtxt{\hat{V}^{\dagger}\hat{V}}},
\end{equation}
with $d$ being the dimension of the matrix (see \cite{Gemmer2004}).
For the random interaction we choose the variance $s^2$ in such a way
that $\eta = 1$ for all interaction matrices coupling adjacent
subunits.

The complete Hamiltonian of the full model system is thus described by
\begin{equation}
  \label{eq:H}
  \hat{H} = \hat{H}_{loc} + \lambda_H \hat{H}_H + \lambda_R \hat{H}_R.
\end{equation}
Because of the normalization of the interaction matrices the numbers
$\lambda_H$ and $\lambda_R$ define the coupling strength between
different sites. The coupling strengths $\lambda_H$ for the Heisenberg
interaction and $\lambda_R$ for the random interaction are chosen so
that $\lambda_R \ll \lambda_H \ll \Delta E$, which is known as the
weak coupling limit.

\begin{figure}
  \centering 
\includegraphics[width=7.5cm]{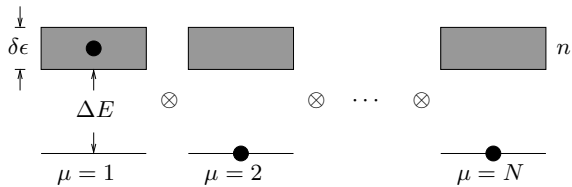}

   \caption{$N$ subunits with ground state and first excitation band
     of width $\delta\varepsilon$ containing $n$ energy levels each.
     Black dots specify the initial states used.}
\label{fig:dingsbums}
\end{figure}

Regardless of the partition scheme chosen (in the $x$ or $z$
direction) each subunit can be seen as a molecule consisting of
several energy bands.  However, the solution for the complete system
is computationally unfeasible for more than a few sites.  If we
restrict ourselves to initial states where only one site is excited
(or superpositions thereof) the Heisenberg interaction does not allow
to leave this subspace of the total Hilbert space.  By choosing also
the random interaction to conserve this subspace we restrict all
further investigations to the single excitation subspace.
Figure~\ref{fig:dingsbums} gives a graphical representation of our
system, with $\delta\varepsilon$ being the width of the first energy
band (all higher excitation bands are neglected here).

Interpreting our model system in terms of a magnetic system, i.e., the
two-level atoms representing coupled spins in a magnetic field for
example, the considered energy transport is equivalent to spin
transport in a gapless system (i.e. $\Delta E = 0$).

%
%
\section{Transport in the $\boldsymbol{z}$ direction}
\label{sec:3}

\subsection{Partitioning scheme}
\label{sec:3:A}

Let us consider the transport perpendicular to the Heisenberg chains
(in the $z$ direction) first.  Then, the partitioning into subunits
yields the following mesoscopic Hamiltonian consisting of a local and
an interaction part
\begin{align}
  \label{eq:hamiltonian}
  \hat{H} &= \h{L} + \h{I}\notag\\
          &= \sum_{\mu = 1}^{N}\h{L}(\mu) + \sum_{\mu = 1}^{N-1}\h{I}(\mu, \mu+1).
\end{align}%
Here $\h{L}(\mu)$ of subunit $\mu$ consists of the constant local
energy splitting, the Heisenberg interaction, and the internal random
couplings of each subunit [cf.\ gray planes in Fig~\ref{fig:3d}(a)].
Since $\lambda_R \ll \lambda_H$ the effect of the internal random
couplings on the spectrum of $\h{L}(\mu)$ may be neglected.  Therefore
the bandwidth $\delta\varepsilon$ is determined by the Heisenberg
interaction given by
\begin{equation}
  \label{eq:dE}
  \delta\varepsilon = 8\lambda_H.
\end{equation}
The last term in Eq.~(\ref{eq:hamiltonian}), $\h{I}(\mu,\mu+1)$,
denotes the interaction between the subunits which is purely random
here, i.e., contains parts of the random interaction Hamiltonian
$\h{R}$ only.

\subsection{Derivation of the TCL master equation}
\label{sec:3:B}

The correlated projection superoperator $\mathcal{P}$ introduced in
Sec.~\ref{sec:1} is of the type as suggested by Breuer
\cite{Breuer2007} and reads
\begin{equation}
  \label{eq:superop}
  \mathcal{P}\dop =
  \sum_{\mu}\trtxt{\PI{\mu}\dop}{}\frac{1}{n}\PI{\mu}\equiv\sum_{\mu}P_{\mu}\frac{1}{n}\PI{\mu},
\end{equation}
with $\PI{\mu}$ being the standard projection operators
\begin{equation}
  \label{eq:projector}
  \PI{\mu} = \sum_{n_{\mu}} \ket{n_{\mu}}\bra{n_{\mu}},
\end{equation}
and $\ket{n_{\mu}}$ the eigenstate of $\h{L}(\mu)$ in the one-particle
excitation subspace, i.e.\ the states in the band of subunit $\mu$
(cf.\ Fig.~\ref{fig:dingsbums}).  Consequently, the number $P_{\mu}$
is just the excitation probability of subunit $\mu$. This choice of
$\mathcal{P}$ thus implements the partitioning scheme required for
studying transport behavior.

Switching to the interaction picture, plugging both the Hamiltonian
(\ref{eq:hamiltonian}) and the projection (\ref{eq:superop}) into
Eq.~(\ref{eq:tcl}) we get
\begin{equation}
  \label{eq:tcl2}
  \dot{P}_{\mu} = -\frac{\lambda_R^2}{n\hbar^2} \sum_{\nu}\int_0^t \D t_1
	\trtxt{\PI{\mu}\comutxt{\h{R}(t)}{\comutxt{\h{R}(t_1)}{\PI{\nu}}}}{}P_{\nu}
\end{equation}
for the second order TCL expansion.  The time dependencies of the
coupling operators refer to the transformation into the interaction
picture and are defined as
\begin{equation}
	\label{eq:mm4}
	\h{R}(t) = \E^{\I \h{L}t}\, \h{R}\, \E^{-\I \h{L}t}.
\end{equation}

By exploiting that $\PI{\mu}$ projects onto eigenstates of
$\h{L}(\mu)$ we can evaluate the trace by using the block structure of
the interaction $\h{I}(\mu,\mu+1)$ between adjacent subunits (see
\cite{Michel2005,Michel2006}), resulting in
\begin{align}
	\label{eq:rate}
	\dod{{P}_{\mu}}{t} = - \gamma_{\mu} \big(2 P_{\mu} - P_{\mu+1} - P_{\mu-1} \big)
\end{align}
with the decay rate
\begin{align}
	\label{eq:dsum}
	\gamma_{\mu} = \frac{2\lambda_R^2}{n\hbar^2}
        \sum_{k,l}^n |\bra{k_{\mu}}\h{R}\ket{l_{\mu+1}}|^2\,
        \frac{\sin(\omega_{kl} t)}{\omega_{kl}}.
\end{align}
The frequency $\omega_{kl}$ refers to the transition between the
eigenstates $k$, $l$.  Equation (\ref{eq:rate}) is basically a rate
equation for the probabilities to find an excitation in subunit $\mu$.

\subsection{Decay rate}
\label{sec:3:C}

Since the interaction between two adjacent subunits is a random matrix
with the above described properties, all matrix elements are
approximately of the same size.  That means that the rate does not
depend on the subunit $\mu$ ($\gamma_{\mu}=\gamma$).  Furthermore, we
can assume $|\bra{k_{\mu}}\h{R}\ket{l_{\mu+1}}|^2 \approx 1$, finding
\begin{equation}
	\label{eq:mm2}
	\gamma =  \frac{2\lambda_R^2}{n\hbar^2} \sum_{k,l} \frac{\sin \omega_{kl} t}{\omega_{kl}}.
\end{equation}
In the following the double sum is treated analogous to the derivation
of Fermi's Golden Rule.

Since the sine cardinal (sinc) of Eq.~(\ref{eq:mm2}) is a
representation of the Dirac $\delta$-distribution
\begin{equation}
	\pi\delta_t(\omega_{kl}) = \lim_{t\rightarrow\infty}\frac{\sin(\omega_{kl}t)}{\omega_{kl}},
\end{equation}
we may approximate the rate for not too small $t$ by
\begin{align}
  \gamma &\approx \frac{2\pi\lambda_{R}^2}{n\hbar}\sum_{k,l}\delta(E_k-E_l).
\end{align}

Replacing the double sum over integrals in the energy space we arrive
at
\begin{equation}
  \gamma \approx \frac{2\pi\lambda_{R}^2}{n \hbar}\int_0^{\delta\varepsilon} g^2(E)\,\D E
\end{equation}
with the state density $g(E)$, i.e., the integral over the square of
the density of states. Since we have neglected the internal random
interaction completely, the state density of the first excitation
subspace is just given by the state density of a Heisenberg spin chain
\begin{equation}
  g(E) = \frac{2n}{\pi\delta\varepsilon}\frac{1}{\sqrt{1-\left(\frac{2E}{\delta\varepsilon}-1\right)^2}}.
\end{equation}
Unfortunately, this function is not square integrable due to
singularities at the boundaries of the spectrum.
However, due to symmetry we have
\begin{equation}
  \int_0^{\delta\varepsilon} g^2(E)\,\D E = 2\int_0^{\delta\varepsilon/2} g^2(E)\,\D E.
\end{equation}
In order to avoid the singularity at $E=0$ we renormalize the number of states in the band.
We introduce the regularized integral
\begin{align}
  \label{eq:fl}
  F_{\Lambda}(n) &= 2\int_{\Lambda}^{\delta\varepsilon/2}\alpha^2 g^2(E)\,\D E \notag\\
                 &= 2\int_{\Lambda}^{\delta\varepsilon/2} \frac{\alpha^2 n^2}{\pi^2 E (\delta\varepsilon-E)}\,\D E,
\end{align}
with $\alpha$ being the factor that renormalizes the number of states.
We assume that for a band consisting of only a few levels $\tilde{n}$
(but still enough to define a density of states), the density of
states is approximately constant.  For a constant density of states
$\tilde{g}(E)$ we simply have
\begin{equation}
  2\int_0^{\delta\varepsilon/2}\tilde{g}^2(E)\,\D E = \frac{\tilde{n}^2}{\delta \varepsilon},
\end{equation}
therefore our renormalization prescription is given by
\begin{equation}
  F_{\Lambda}(\tilde{n}) = \frac{\tilde{n}^2}{\delta\varepsilon}.
\end{equation}
Using this result to solve Eq.~(\ref{eq:fl}) for $\alpha$ at constant $\tilde{n}$ yields
\begin{equation}
  \alpha = \frac{\pi}{\sqrt{2\ln(\delta\varepsilon/\Lambda - 1)}}.
\end{equation}
This allows us to calculate the physical limit of the renormalization procedure, i.e.,
\begin{equation}
  \lim\limits_{\Lambda \rightarrow 0} F_{\Lambda}(n) = \frac{n^2}{\delta\varepsilon}
\end{equation}
which is the same value as for a constant density of states.
This finally leads to the relaxation rate
\begin{equation}
  \label{eq:gamma}
  \gamma = \frac{2\pi\lambda_R^2 n}{\hbar \delta\varepsilon}.
\end{equation}
The approximation introduced by Fermi's Golden Rule is only valid in
the linear regime (see \cite{Gemmer2004}), i.e.,
\begin{equation}
  \label{eq:peter2}
  \frac{4\pi^2 n \lambda_R^2}{\delta\varepsilon^2} \ll 1.
\end{equation}

\subsection{Solution of the TCL master equation}
\label{sec:3:D}

\begin{figure}
  \centering 
  \includegraphics{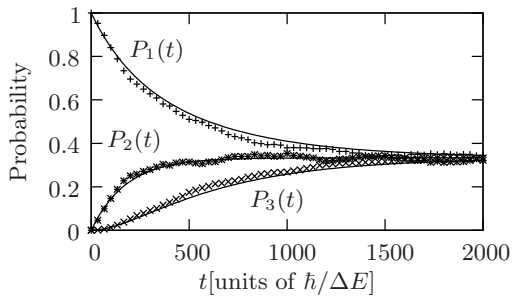}
  \caption{Perpendicular transport: probability to find the excitation
    in subunit $\mu=1,2,3$.  Comparison of the numerical solution of
    the Schr\"odinger equation (crosses) and second-order TCL (lines).
    ($N = 3$, $n = 600$, $\lambda_R = 5\cdot 10^{-4}\Delta E$,
    $\lambda_H = 6.25\cdot 10^{-2} \Delta E$)}
  \label{fig:normal}
\end{figure}

Figure \ref{fig:normal} shows both the numerical results for the
solution of the full Schr\"odinger equation and the solution of the
rate equation (\ref{eq:rate}), according to the above derived approximation for the rate $\gamma$ [cf.~Eq.~(\ref{eq:gamma})].
Both are in reasonably good agreement.

Equation (\ref{eq:rate}) is a discrete version of the diffusion
equation, which does not change when regarding the thermodynamic limit
($n, N \rightarrow \infty$, $n\lambda_R^2=\textup{const}$). For a
$\delta$-shaped excitation at $t = 0$ its solution is a Gaussian
function, the variance of which grows linear in time. Therefore, it is
evident that the heat transport is normal perpendicular to the chains.

%
%
\section{Transport in the $\boldsymbol{x}$ direction}
\label{sec:4}

\subsection{Partitioning scheme}
\label{sec:4:A}

In the following let us concentrate on the transport in the $x$
direction, i.e., parallel to the chains.  Thus, we have a slightly
different partition of the total Hamiltonian.  Besides the local
energy splitting, the local part $\h{L}$ of the mesoscopic Hamiltonian
(\ref{eq:hamiltonian}) contains random interactions only:
\begin{equation}
  \label{eq:mm3}
  \h{L} = \sum_{\mu = 1}^{N}\big[\h{loc}(\mu) + \lambda_R\h{R}(\mu)\big].
\end{equation}
In contrast, the interaction between the subunits consists of a
Heisenberg and a random part,
\begin{equation}
  \label{eq:hamiltonian2}
  \h{I} = \sum_{\mu = 1}^{N-1}\big[\lambda_H\h{H}(\mu,\mu+1) + \lambda_R\h{R}(\mu,\mu+1)\big].
\end{equation}%

In the one-particle excitation subspace the commutator relations
\begin{equation}
  \label{eq:commu}
  \comutxt{\h{H}}{\h{L}} = \comutxt{\h{H}}{\h{R}} = 0
\end{equation}
are satisfied.  If the dynamics induced by the local part $\h{L}$ and
the Heisenberg $\h{H}$ is absorbed in the transformation into the
interaction picture, the random part of the interaction transforms
into
\begin{align}
  \label{eq:ibild}
  \h{R}(t) &= \E^{\I\left(\h{H} + \h{L}\right)t}\, \h{R}\, \E^{-\I\left(\h{H} + \h{L}\right)t}\notag\\
           &= \E^{\I \h{L}t}\, \h{R}\, \E^{-\I\h{L}t},
\end{align}
where Eq.~(\ref{eq:commu}) has been used. Note that this is not the
standard interaction picture as used above, but a special one allowing
us to treat the transport in the $x$ direction in a similar manner as
in the $z$ direction.  According to this transformation the derivation
of the second order TCL master equation in Sec.~\ref{sec:3:B},
especially Eqs.~(\ref{eq:tcl2}), (\ref{eq:rate}) and (\ref{eq:dsum}),
remain unchanged.

\subsection{Decay rate}
\label{sec:4:B}

However, the computation of the rate (\ref{eq:mm2}) is different here.
For calculating the local band structure we consider just a random
matrix of dimension $n$, drawn from a Gaussian unitary ensemble. From
random matrix theory \cite{Mehta1991} it is known that the density of
levels $\zeta(x)$ for such a random Hermitian matrix consisting of
elements with zero mean and unit variance for both real and imaginary
parts is given by
\begin{equation}
  \zeta(x) = \frac{1}{\pi}\sqrt{2n-x^2}.
\end{equation}
Mapping this to a density of energy levels leads to
\begin{equation}
  \label{eq:mm5}
  g(E) = \frac{8n}{\pi\delta\varepsilon}\sqrt{\frac{\delta\varepsilon^2}{4}-E^2}.
\end{equation}
We rescale the variance to the interaction strength $\lambda_R$, which
gives for the bandwidth \
\begin{equation}
  \label{eq:band}
  \delta\varepsilon = 4\sqrt{n}\lambda_R.
\end{equation}

\begin{figure}
  \centering 
  \includegraphics{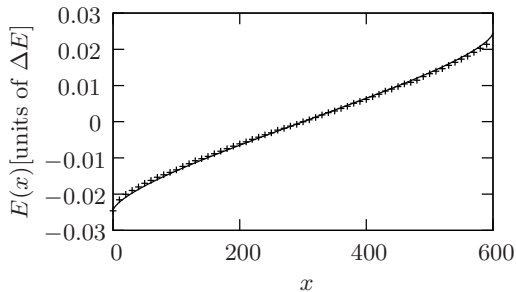}
  \caption{Comparison of the eigenvalues of $\h{L}$ and a random
    matrix drawn from a Gaussian unitary ensemble. ($n=600$,
    $\lambda_R = 5\cdot10^{-4}\Delta E$)}
\label{fig:gue}  
\end{figure}

In order to check whether our local Hamiltonian $\h{L}$ can be
approximated by such a random matrix, we compare the eigenvalues
$E(x)$ of both matrices. Using
\begin{equation}
  \dod{E}{x} = \frac{1}{g[E(x)]}
\end{equation}
and separating variables yields
\begin{equation}
  \frac{8n}{\pi\delta\varepsilon}\sqrt{\frac{\delta\varepsilon^2}{4}-E^2}\,\D E = \D x,
\end{equation}
with the state density (\ref{eq:mm5}). This expression cannot be
solved analytically for $E$, so we compare the numerical solution for
discrete values of $x$ with the eigenvalues of $\h{L}$. As
Fig.~\ref{fig:gue} shows, $\h{L}$ may indeed be approximated by a
random matrix drawn from a Gaussian unitary ensemble.  However, by
plugging Eq.~(\ref{eq:band}) into Eq.~(\ref{eq:peter2}) one gets a
constant value of $\pi^2/4$ which is definitely not small compared to
one.  Thus, the requirement for the linear regime is violated and the
derivation of the rate according to Fermi's Golden Rule can no longer
be applied.

The approximation used in Sec.~\ref{sec:3:B} is analogous to Fermi's
Golden Rule.  All transitions in Eq.~(\ref{eq:mm2}) from $l$ to $k$
are weighted by the respective value of the sinc function which
changes its shape for increasing times to approach a delta peak for
$t\rightarrow\infty$.  In the situation described above the decay
takes place within the linear regime, i.e., at an intermediate time
scale.  That means that all possible transition frequencies are
distributed below the peak.  Thus the sum in Eq.~(\ref{eq:mm2}) can be
approximated by the area under the peak (see \cite{Gemmer2004}).

This is not the case here.  The decay happens on a much shorter time
scale, when the peak is extremely broad.  Therefore almost any
transition frequency belongs to the maximum of the peak.  Thus the
sinc in Eq.~(\ref{eq:mm2}) should better be approximated by the
maximum value of the peak, instead of the area under the peak.  The
maximum value grows with time according to $t$. Thus the double sum
over sinc functions could be approximated by $n^2 t$.  This means that
we get the relaxation rate
\begin{equation}
  \label{eq:gammat}
  \gamma = \frac{2 n \lambda_R^2}{\hbar^2}\, t.
\end{equation}

\subsection{Solution of the TCL master equation}
\label{sec:4:C}

The solution of Eq.~(\ref{eq:rate}) with the diffusion coefficient
(\ref{eq:gammat}) defines the occupation probabilities in the
interaction picture $P_{\mu}^{\textup{int}}$.  Note that in the other
direction the occupation probabilities of the interaction picture have
been equivalent to the occupation probabilities in the Schr\"odinger
picture.  This is not the case for the present situation because of
the special choice of the interaction picture.  Remember that we have
used not only the local Hamiltonian for the transformation into the
interaction picture, but also a part of the inter-subsystem
interaction [cf.~Eq.~(\ref{eq:ibild})].

Since we are interested in the occupation probabilities in the Schr\"odinger picture $P_{\mu}^{\textup{s}}$ we need to calculate the inverse transformation of the density operator
\begin{equation}
  \label{eq:back}
  \mathcal{P}\dop^{\textup{s}} = \E^{-\I\h{H}t}\mathcal{P}\dop^{\textup{int}}\E^{\I\h{H}t},
\end{equation}
where the diagonal elements $\mathcal{P}\dop^{\textup{s}}_{\mu\mu}$ are the occupation probabilities $P_{\mu}^{\textup{s}}$.
The off-diagonal elements of $\mathcal{P}\dop^{\textup{int}}$ can be
computed by replacing the projector (\ref{eq:projector}) with another
one projecting out off-diagonal elements as well. The dynamics of the
diagonal and the off-diagonal elements decouple so that diagonal
initial states remain diagonal for all time.

\begin{figure}
  \centering 
  
  \includegraphics{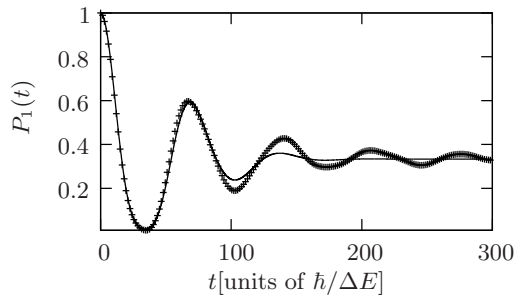}
  \caption{Parallel transport: probability to find the excitation in
    the first subunit ($\mu=1$).  Comparison of the numerical solution
    of the Schr\"odinger equation (crosses) and second-order TCL
    (solid line). (Same parameters as for Fig.~\ref{fig:normal})}
\label{fig:ball}  
\end{figure}

Thus using Eq.~(\ref{eq:back}) for the inverse transformation we get
the time-dependent solution of the probabilities in the Schr\"odinger
picture.  In Fig.~\ref{fig:ball} the numerical solution of the
Schr\"odinger equation is compared with the TCL prediction. Again,
there is a very good agreement between the exact solution and our
second order approximation.

\subsection{Spatial variance}
\label{sec:4:D}

\begin{figure}
  \centering 
  
  \includegraphics{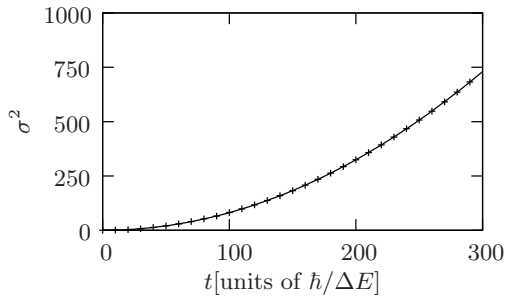}
  \caption{Variance of an excitation initially at subunit $\mu_0 =
    150$.  Second-order TCL prediction (crosses) and quadratic fit
    (solid line).  ($N = 300$, $n = 600$, $\lambda_R = 5\cdot
    10^{-4}$, $\lambda_H~=~6.25~\cdot~10^{-2}$)}
\label{fig:var}  
\end{figure}

To classify the transport behavior in the $x$ direction a very large
system has to be considered, so that the initial excitation does not
reach the boundaries of the system during the relaxation time.  Since
the solution of the time-dependent Schr\"odinger equation becomes
unfeasible the second-order TCL prediction has been used for
subsequent numerical integration. The variance of an excitation
initially at $\mu = \mu_0$,
\begin{equation}
  \label{eq:var}
  \sigma^2(t) = \sum\limits_{\mu=1}^{N}P_{\mu}^{(s)}(t)(\mu - \mu_0)^2,
\end{equation}
shown in Fig.~\ref{fig:var} grows quadratic in time, i.e., the
transport is ballistic. Here we have considered a system with $N =
300$ subunits and an initial excitation at $\mu_0 = 150$ solving the
TCL master equation. This is also valid in the thermodynamic limit as
$\gamma(t)$ does not change. Numerical investigations show that the
transport behavior is largely independent of $\gamma(t)$. Ballistic
transport is observed as long as $\lambda_H t \gg \gamma(t)$ on all
relevant time-scales.

\section{Conclusions}

In the present paper we have demonstrated how the abstract method of
correlated projection superoperators for the TCL master equation
\cite{Breuer2006,Breuer2007} can be used to analyze the transport
behavior of a three-dimensional solid state model: a system of coupled
two-level atoms with an anisotropic interaction.  The analysis is
based on the following preconditions:
\begin{enumerate}
\item a \emph{partitioning scheme} in position space to consider the
  transport in one direction of the model, thus introducing a
  projection superoperator
\item the \emph{convergence} of the TCL expansion in second order (a
  wrong projection superoperator leads to a diverging expansion, or
  large higher than second orders)
\item an \emph{approximation scheme} for computing the decay rate to
  avoid numerical integration
\end{enumerate}
According to those central points a reduced dynamical description of
the complex quantum model is derived which can be analyzed, e.g., to
classify the transport behavior of the system.

By a comparison of the TCL prediction with the exact numerical
solution of the complete Schr\"odinger equation of our model system we
have shown that the results of the method are in very good accordance
with the real dynamical behavior of the system.  Having established a
method which efficiently describes the dynamical properties of a
complex quantum model the transport behavior can be classified by
either an analytic analysis of the solution of the reduced dynamical
equations or by a numerical investigation.  Here, the simplicity of
the reduced equations in comparison to the exact system of
differential equations allows us to investigate the dynamical
properties of a much larger system which is not accessible to a direct
investigation.

The analysis shows that the model features two very different types of
transport behavior in the $x$ and $z$ directions, perpendicular or
parallel to the chains, respectively.  In the $z$ direction we have
found a standard statistical decay behavior following a diffusion
equation on the basis of the mesoscopic subunits. In this way
diffusive behavior has been derived from first principles on a
mesoscopic scale whereas the dynamics on the microscopic scale (i.e.,
of a single spin) is obviously non-diffusive. This indicates that the
transport behavior is not only a property of a system \emph{per se},
but also depends on the way we are looking at it.  In contrast the
model shows ballistic behavior parallel to the chains which is
demonstrated by the features of the reduced dynamical equations. Note
that this behavior is similar to investigations of large anisotropies
within the heat conductivity of cuprates
\cite{Sologubenko2000,Hess2001}.

In conclusion this method of a correlated projection superoperators
within TCL allows to investigate the dynamical behavior of 3D model
systems on a mesoscopic scale.  It is useful both in the case of a
statistical decay according to a diffusion equation and the ballistic
case, where time dependent rates are important.

\begin{acknowledgments}
We thank H.-P.\ Breuer, M.\ Henrich, F.\ Rempp, G.\ Reuther, H.\
Schmidt, H.\ Schr\"oder, J.\ Teifel and P.\ Vidal for fruitful
discussions. Financial support by the Deutsche Forschungsgemeinschaft
is gratefully acknowledged.
\end{acknowledgments}


\begin{thebibliography}{28}
\expandafter\ifx\csname natexlab\endcsname\relax\def\natexlab#1{#1}\fi
\expandafter\ifx\csname bibnamefont\endcsname\relax
  \def\bibnamefont#1{#1}\fi
\expandafter\ifx\csname bibfnamefont\endcsname\relax
  \def\bibfnamefont#1{#1}\fi
\expandafter\ifx\csname citenamefont\endcsname\relax
  \def\citenamefont#1{#1}\fi
\expandafter\ifx\csname url\endcsname\relax
  \def\url#1{\texttt{#1}}\fi
\expandafter\ifx\csname urlprefix\endcsname\relax\def\urlprefix{URL }\fi
\providecommand{\bibinfo}[2]{#2}
\providecommand{\eprint}[2][]{\url{#2}}

\bibitem[{\citenamefont{Zotos et~al.}(1997)\citenamefont{Zotos, Naef, and
  Prelovsek}}]{Zotos1997}
\bibinfo{author}{\bibfnamefont{X.}~\bibnamefont{Zotos}},
  \bibinfo{author}{\bibfnamefont{F.}~\bibnamefont{Naef}}, \bibnamefont{and}
  \bibinfo{author}{\bibfnamefont{P.}~\bibnamefont{Prelovsek}},
  \bibinfo{journal}{Phys. Rev. B} \textbf{\bibinfo{volume}{55}},
  \bibinfo{pages}{11029} (\bibinfo{year}{1997}).

\bibitem[{\citenamefont{Prosen}(1999)}]{Prosen1999}
\bibinfo{author}{\bibfnamefont{T.}~\bibnamefont{Prosen}},
  \bibinfo{journal}{Phys. Rev. E} \textbf{\bibinfo{volume}{60}},
  \bibinfo{pages}{3949} (\bibinfo{year}{1999}).

\bibitem[{\citenamefont{Kl\"umper and Sakai}(2002)}]{Kluemper2002}
\bibinfo{author}{\bibfnamefont{A.}~\bibnamefont{Kl\"umper}} \bibnamefont{and}
  \bibinfo{author}{\bibfnamefont{K.}~\bibnamefont{Sakai}}, \bibinfo{journal}{J.
  Phys. A} \textbf{\bibinfo{volume}{35}}, \bibinfo{pages}{2173}
  (\bibinfo{year}{2002}).

\bibitem[{\citenamefont{Heidrich-Meisner
  et~al.}(2003)\citenamefont{Heidrich-Meisner, Honecker, Cabra, and
  Brenig}}]{Heidrich-Meisner2003}
\bibinfo{author}{\bibfnamefont{F.}~\bibnamefont{Heidrich-Meisner}},
  \bibinfo{author}{\bibfnamefont{A.}~\bibnamefont{Honecker}},
  \bibinfo{author}{\bibfnamefont{D.~C.} \bibnamefont{Cabra}}, \bibnamefont{and}
  \bibinfo{author}{\bibfnamefont{W.}~\bibnamefont{Brenig}},
  \bibinfo{journal}{Phys. Rev. B} \textbf{\bibinfo{volume}{68}},
  \bibinfo{eid}{134436} (\bibinfo{year}{2003}).

\bibitem[{\citenamefont{Saito}(2003{\natexlab{a}})}]{Saito2003}
\bibinfo{author}{\bibfnamefont{K.}~\bibnamefont{Saito}},
  \bibinfo{journal}{Phys. Rev. B} \textbf{\bibinfo{volume}{67}},
  \bibinfo{eid}{064410} (\bibinfo{year}{2003}{\natexlab{a}}).

\bibitem[{\citenamefont{Jung et~al.}(2006)\citenamefont{Jung, Helmes, and
  Rosch}}]{Jung2006}
\bibinfo{author}{\bibfnamefont{P.}~\bibnamefont{Jung}},
  \bibinfo{author}{\bibfnamefont{R.~W.} \bibnamefont{Helmes}},
  \bibnamefont{and} \bibinfo{author}{\bibfnamefont{A.}~\bibnamefont{Rosch}},
  \bibinfo{journal}{Phys. Rev. Lett.} \textbf{\bibinfo{volume}{96}},
  \bibinfo{eid}{067202} (\bibinfo{year}{2006}).

\bibitem[{\citenamefont{Kubo}(1957)}]{Kubo1957}
\bibinfo{author}{\bibfnamefont{R.}~\bibnamefont{Kubo}}, \bibinfo{journal}{J.
  Phys. Soc. Jpn.} \textbf{\bibinfo{volume}{12}}, \bibinfo{pages}{570}
  (\bibinfo{year}{1957}).

\bibitem[{\citenamefont{Kubo et~al.}(1991)\citenamefont{Kubo, Toda, and
  Hashitsume}}]{Kubo1991}
\bibinfo{author}{\bibfnamefont{R.}~\bibnamefont{Kubo}},
  \bibinfo{author}{\bibfnamefont{M.}~\bibnamefont{Toda}}, \bibnamefont{and}
  \bibinfo{author}{\bibfnamefont{N.}~\bibnamefont{Hashitsume}},
  \emph{\bibinfo{title}{Statistical {P}hysics {II}: {N}onequilibrium
  {S}tatistical {M}echanics}}, no.~\bibinfo{number}{31} in
  \bibinfo{series}{Solid-State Sciences} (\bibinfo{publisher}{Springer},
  \bibinfo{address}{Berlin, Heidelberg, New-York}, \bibinfo{year}{1991}),
  \bibinfo{edition}{2nd} ed.

\bibitem[{\citenamefont{Luttinger}(1964)}]{Luttinger1964}
\bibinfo{author}{\bibfnamefont{J.~M.} \bibnamefont{Luttinger}},
  \bibinfo{journal}{Phys. Rev.} \textbf{\bibinfo{volume}{135}},
  \bibinfo{pages}{A1505} (\bibinfo{year}{1964}).

\bibitem[{\citenamefont{Gemmer et~al.}(2006)\citenamefont{Gemmer, Steinigeweg,
  and Michel}}]{Gemmer2006}
\bibinfo{author}{\bibfnamefont{J.}~\bibnamefont{Gemmer}},
  \bibinfo{author}{\bibfnamefont{R.}~\bibnamefont{Steinigeweg}},
  \bibnamefont{and} \bibinfo{author}{\bibfnamefont{M.}~\bibnamefont{Michel}},
  \bibinfo{journal}{Phys. Rev. B} \textbf{\bibinfo{volume}{73}},
  \bibinfo{eid}{104302} (\bibinfo{year}{2006}).

\bibitem[{\citenamefont{Gobert et~al.}(2005)\citenamefont{Gobert, Kollath,
  Schollw\"ock, and Sch\"utz}}]{Gobert2005}
\bibinfo{author}{\bibfnamefont{D.}~\bibnamefont{Gobert}},
  \bibinfo{author}{\bibfnamefont{C.}~\bibnamefont{Kollath}},
  \bibinfo{author}{\bibfnamefont{U.}~\bibnamefont{Schollw\"ock}},
  \bibnamefont{and} \bibinfo{author}{\bibfnamefont{G.}~\bibnamefont{Sch\"utz}},
  \bibinfo{journal}{Phys. Rev. E} \textbf{\bibinfo{volume}{71}},
  \bibinfo{pages}{036102} (\bibinfo{year}{2005}).

\bibitem[{\citenamefont{Mej\'{\i}a-Monasterio
  et~al.}(2005)\citenamefont{Mej\'{\i}a-Monasterio, Prosen, and
  Casati}}]{Mejia-Monasterio2005}
\bibinfo{author}{\bibfnamefont{C.}~\bibnamefont{Mej\'{\i}a-Monasterio}},
  \bibinfo{author}{\bibfnamefont{T.}~\bibnamefont{Prosen}}, \bibnamefont{and}
  \bibinfo{author}{\bibfnamefont{G.}~\bibnamefont{Casati}},
  \bibinfo{journal}{Europhys. Lett.} \textbf{\bibinfo{volume}{72}},
  \bibinfo{pages}{520} (\bibinfo{year}{2005}).

\bibitem[{\citenamefont{Steinigeweg et~al.}(2006)\citenamefont{Steinigeweg,
  Gemmer, and Michel}}]{Steinigeweg2006}
\bibinfo{author}{\bibfnamefont{R.}~\bibnamefont{Steinigeweg}},
  \bibinfo{author}{\bibfnamefont{J.}~\bibnamefont{Gemmer}}, \bibnamefont{and}
  \bibinfo{author}{\bibfnamefont{M.}~\bibnamefont{Michel}},
  \bibinfo{journal}{Europhys. Lett.} \textbf{\bibinfo{volume}{75}},
  \bibinfo{pages}{406} (\bibinfo{year}{2006}).

\bibitem[{\citenamefont{Saito}(2003{\natexlab{b}})}]{Saito2003a}
\bibinfo{author}{\bibfnamefont{K.}~\bibnamefont{Saito}},
  \bibinfo{journal}{Europhys. Lett.} \textbf{\bibinfo{volume}{61}},
  \bibinfo{pages}{34} (\bibinfo{year}{2003}{\natexlab{b}}).

\bibitem[{\citenamefont{Michel et~al.}(2003)\citenamefont{Michel, Hartmann,
  Gemmer, and Mahler}}]{Michel2003}
\bibinfo{author}{\bibfnamefont{M.}~\bibnamefont{Michel}},
  \bibinfo{author}{\bibfnamefont{M.}~\bibnamefont{Hartmann}},
  \bibinfo{author}{\bibfnamefont{J.}~\bibnamefont{Gemmer}}, \bibnamefont{and}
  \bibinfo{author}{\bibfnamefont{G.}~\bibnamefont{Mahler}},
  \bibinfo{journal}{Eur. Phys. J. B} \textbf{\bibinfo{volume}{34}},
  \bibinfo{pages}{325} (\bibinfo{year}{2003}).

\bibitem[{\citenamefont{Breuer and Petruccione}(2002)}]{Breuer2002}
\bibinfo{author}{\bibfnamefont{H.-P.} \bibnamefont{Breuer}} \bibnamefont{and}
  \bibinfo{author}{\bibfnamefont{F.}~\bibnamefont{Petruccione}},
  \emph{\bibinfo{title}{The {T}heory of {O}pen {Q}uantum {S}ystems}}
  (\bibinfo{publisher}{Oxford University Press}, \bibinfo{address}{Oxford},
  \bibinfo{year}{2002}).

\bibitem[{\citenamefont{Michel et~al.}(2004)\citenamefont{Michel, Gemmer, and
  Mahler}}]{Michel2004}
\bibinfo{author}{\bibfnamefont{M.}~\bibnamefont{Michel}},
  \bibinfo{author}{\bibfnamefont{J.}~\bibnamefont{Gemmer}}, \bibnamefont{and}
  \bibinfo{author}{\bibfnamefont{G.}~\bibnamefont{Mahler}},
  \bibinfo{journal}{Eur. Phys. J. B} \textbf{\bibinfo{volume}{42}},
  \bibinfo{pages}{555} (\bibinfo{year}{2004}).

\bibitem[{\citenamefont{Gemmer et~al.}(2004)\citenamefont{Gemmer, Michel, and
  Mahler}}]{Gemmer2004}
\bibinfo{author}{\bibfnamefont{J.}~\bibnamefont{Gemmer}},
  \bibinfo{author}{\bibfnamefont{M.}~\bibnamefont{Michel}}, \bibnamefont{and}
  \bibinfo{author}{\bibfnamefont{G.}~\bibnamefont{Mahler}},
  \emph{\bibinfo{title}{Quantum {T}hermodynamics}}, \bibinfo{number}{Lecture
  Notes in Physics, Vol. 657} (\bibinfo{publisher}{Springer},
  \bibinfo{address}{Berlin}, \bibinfo{year}{2004}).

\bibitem[{\citenamefont{Michel et~al.}(2005)\citenamefont{Michel, Mahler, and
  Gemmer}}]{Michel2005}
\bibinfo{author}{\bibfnamefont{M.}~\bibnamefont{Michel}},
  \bibinfo{author}{\bibfnamefont{G.}~\bibnamefont{Mahler}}, \bibnamefont{and}
  \bibinfo{author}{\bibfnamefont{J.}~\bibnamefont{Gemmer}},
  \bibinfo{journal}{Phys. Rev. Lett.} \textbf{\bibinfo{volume}{95}},
  \bibinfo{eid}{180602} (\bibinfo{year}{2005}).

\bibitem[{\citenamefont{Michel et~al.}(2006)\citenamefont{Michel, Gemmer, and
  Mahler}}]{Michel2006}
\bibinfo{author}{\bibfnamefont{M.}~\bibnamefont{Michel}},
  \bibinfo{author}{\bibfnamefont{J.}~\bibnamefont{Gemmer}}, \bibnamefont{and}
  \bibinfo{author}{\bibfnamefont{G.}~\bibnamefont{Mahler}},
  \bibinfo{journal}{Phys. Rev. E} \textbf{\bibinfo{volume}{73}},
  \bibinfo{eid}{016101} (\bibinfo{year}{2006}).

\bibitem[{\citenamefont{Breuer et~al.}(2006)\citenamefont{Breuer, Gemmer, and
  Michel}}]{Breuer2006}
\bibinfo{author}{\bibfnamefont{H.-P.} \bibnamefont{Breuer}},
  \bibinfo{author}{\bibfnamefont{J.}~\bibnamefont{Gemmer}}, \bibnamefont{and}
  \bibinfo{author}{\bibfnamefont{M.}~\bibnamefont{Michel}},
  \bibinfo{journal}{Phys. Rev. E} \textbf{\bibinfo{volume}{73}},
  \bibinfo{eid}{016139} (\bibinfo{year}{2006}).

\bibitem[{\citenamefont{Breuer}(2007)}]{Breuer2007}
\bibinfo{author}{\bibfnamefont{H.-P.} \bibnamefont{Breuer}},
  \bibinfo{journal}{Phys. Rev. A} \textbf{\bibinfo{volume}{75}},
  \bibinfo{eid}{022103} (\bibinfo{year}{2007}).

\bibitem[{\citenamefont{Nakajima}(1958)}]{Nakajima1958}
\bibinfo{author}{\bibfnamefont{S.}~\bibnamefont{Nakajima}},
  \bibinfo{journal}{Progr. Theo. Phys.} \textbf{\bibinfo{volume}{20}},
  \bibinfo{pages}{948} (\bibinfo{year}{1958}).

\bibitem[{\citenamefont{Zwanzig}(1960)}]{Zwanzig1960}
\bibinfo{author}{\bibfnamefont{R.}~\bibnamefont{Zwanzig}}, \bibinfo{journal}{J.
  Chem. Phys.} \textbf{\bibinfo{volume}{33}}, \bibinfo{pages}{1338}
  (\bibinfo{year}{1960}).

\bibitem[{\citenamefont{Schollw\"ock et~al.}(2004)\citenamefont{Schollw\"ock,
  Richter, Farnell, and Bishop}}]{Schollwock2004}
\bibinfo{editor}{\bibfnamefont{U.}~\bibnamefont{Schollw\"ock}},
  \bibinfo{editor}{\bibfnamefont{J.}~\bibnamefont{Richter}},
  \bibinfo{editor}{\bibfnamefont{D.~J.} \bibnamefont{Farnell}},
  \bibnamefont{and} \bibinfo{editor}{\bibfnamefont{R.~F.}
  \bibnamefont{Bishop}}, eds., \emph{\bibinfo{title}{Quantum {M}agnetism}},
  \bibinfo{number}{Lecture Notes in Physics, Vol. 645}
  (\bibinfo{publisher}{Springer}, \bibinfo{address}{Berlin},
  \bibinfo{year}{2004}).

\bibitem[{\citenamefont{Sologubenko et~al.}(2000)\citenamefont{Sologubenko,
  Giann\'o, Ott, Ammerahl, and Revcolevschi}}]{Sologubenko2000}
\bibinfo{author}{\bibfnamefont{A.~V.} \bibnamefont{Sologubenko}},
  \bibinfo{author}{\bibfnamefont{K.}~\bibnamefont{Giann\'o}},
  \bibinfo{author}{\bibfnamefont{H.~R.} \bibnamefont{Ott}},
  \bibinfo{author}{\bibfnamefont{U.}~\bibnamefont{Ammerahl}}, \bibnamefont{and}
  \bibinfo{author}{\bibfnamefont{A.}~\bibnamefont{Revcolevschi}},
  \bibinfo{journal}{Phys. Rev. Lett.} \textbf{\bibinfo{volume}{84}},
  \bibinfo{pages}{2714} (\bibinfo{year}{2000}).

\bibitem[{\citenamefont{Hess et~al.}(2001)\citenamefont{Hess, Baumann,
  Ammerahl, B\"uchner, Heidrich-Meisner, Brenig, and Revcolevschi}}]{Hess2001}
\bibinfo{author}{\bibfnamefont{C.}~\bibnamefont{Hess}},
  \bibinfo{author}{\bibfnamefont{C.}~\bibnamefont{Baumann}},
  \bibinfo{author}{\bibfnamefont{U.}~\bibnamefont{Ammerahl}},
  \bibinfo{author}{\bibfnamefont{B.}~\bibnamefont{B\"uchner}},
  \bibinfo{author}{\bibfnamefont{F.}~\bibnamefont{Heidrich-Meisner}},
  \bibinfo{author}{\bibfnamefont{W.}~\bibnamefont{Brenig}}, \bibnamefont{and}
  \bibinfo{author}{\bibfnamefont{A.}~\bibnamefont{Revcolevschi}},
  \bibinfo{journal}{Phys. Rev. B} \textbf{\bibinfo{volume}{64}},
  \bibinfo{pages}{184305} (\bibinfo{year}{2001}).

\bibitem[{\citenamefont{Mehta}(1991)}]{Mehta1991}
\bibinfo{author}{\bibfnamefont{M.~L.} \bibnamefont{Mehta}},
  \emph{\bibinfo{title}{Random Matrices}} (\bibinfo{publisher}{Academic Press},
  \bibinfo{address}{Boston}, \bibinfo{year}{1991}).

\end{thebibliography}

\end{document}